\documentclass[showpacs,superscriptaddress,amsmath,amssymb,reprint,prl]{revtex4-1}

\usepackage{graphicx}
\usepackage[table]{xcolor}
\usepackage{bm}

\begin{document}

\title{Shallow Angle Inverse Compton Scattering}

\author{B.H. Schaap}
\email{schaap@physics.ucla.edu}

\author{M. Lenz}

\author{P. Musumeci}

\affiliation{Department of Physics and Astronomy, University of California at Los Angeles, Los Angeles, California 90095, USA}

\date{\today}

\begin{abstract}

Inverse Compton Scattering at shallow angles can be used to tune the spectrum and increase the brightness of the emitted radiation. Here, we report on the demonstration of emission of visible light by a 4.7 MeV $e$-beam crossing a 780 nm laser at 5.8$^{\rm o}$ angle. Angular and spectral measurements of the radiation are in agreement with analytical and numerical predictions. We observe a characteristic dependence on the incoming polarization, not present in conventional head-on scattering, with suppression of the emission for $p$-polarization at condition equivalent to relativistic Brewster reflection.
\end{abstract}


\maketitle 

X-ray sources permit the interrogation of a wide range of materials, including integrated circuits, cells and electrocatalyst \cite{Aidukas2024High-performancePtychography,Egawa2024ObservationLaser,Li2025OperandoCells}. 
Unfortunately, high brightness sources in the X-ray spectral range are limited to large scale facilities such as synchrotrons and free electron lasers \cite{Wang2024MillijouleLaser,Yan2024Terawatt-attosecondRate,Franz2024Terawatt-scaleLaser}.  Further routes towards compact and powerful X-ray sources are therefore a subject of active research. A promising approach is Inverse Compton Scattering (ICS) radiation from relativistic electrons colliding with a laser pulse \cite{Fiocco1963ThomsonBeam,ride1995thomson,Leemans1996X-RayScattering,albert2010characterization,Powers2013Quasi-monoenergeticSource,sakai2024hard}. 
Due to the short wavelength of the laser wave, X-ray emission can be attained using modest electron beam energies enabling the use small scale accelerators. This has led to the development of several compact X-ray sources based on head-on inverse Compton scattering \cite{Graves2014CompactKHz,Deitrick2018High-brillianceSource,petrillo2023state,Alkadi2025CommissioningX-rays/s}.  The drawback of the low $e$-beam energy required for head-on scattering, where the laser pulse is counter-propagating, is the large opening angle of the emission (inversely proportional to the electron Lorentz factor $\gamma$)\cite{Einstein1905ZurKorper}, which together with the small cross-section \cite{Klein1928The2} significantly restricts the source brightness compared to large scale facilities.



The crossing angle $\theta_0$ of the laser pulse with the $e$-beam allows for additional tuning of the radiation wavelength, which is given by  \cite{Debus2010Traveling-waveSources}
\begin{equation} \label{eq: resonance condition}
    \lambda_{\rm rad} = \frac{\lambda_0}{2\gamma^2 \left(1 - \beta \cos\theta_0\right)} \left( 1+ \gamma^2\theta^2 + \frac{a_0^2}{2}\right),
\end{equation}
where $\lambda_0$ is the wavelength of the laser beam, $\gamma$ the Lorentz factor, $\beta$ is the velocity of the electrons normalized to the speed of light, $\theta$ is the scattering angle from the $e$-beam axis and $a_0$ the normalized vector potential of the laser beam. Here and in the following, we assume that the electron beam is relativistic ($\gamma \gg 1$) and redshifting from ponderomotive forces is negligible ($a_0\ll1$). Equation (\ref{eq: resonance condition}) indicates that higher $e$-beam energy can be employed than minimally required for head-on scattering with $\theta_0=  180^{o}$. For example, by changing from head-on to near co-propagating geometry with a shallow crossing angle of $\theta_0 = 4.4^{o}$ for a 780 nm laser, in order to tune
the radiation wavelength to $\lambda_{\rm rad} = 3.3$ nm in the center of the water window \cite{Egawa2024ObservationLaser}, the $e$-beam energy should be increased from 4 MeV to 100 MeV -still attainable in compact accelerator setups \cite{Faillace2022HighApplications,Fuchs2009Laser-drivenSource}.

In addition to the brightness increase due to relativistic beaming, there are many other advantages of employing an $e$-beam with higher beam energy. Firstly, space charge forces within the $e$-beam are reduced significantly allowing for tighter transverse focusing and improved compression leading to bright, ultrashort radiation pulses \cite{Chao2023HandbookEngineering}. Secondly, it unfolds the potential of operating in a superradiant regime: microbunching methods required for superradiance, either produce better results \cite{Schaap2023PonderomotiveSource} or exclusively operate at higher beam energy  \cite{Liu1998ExperimentalAcceleration,Duris2012InverseSources,Sudar2018DemonstrationBeam} and the superradiant photon yield scales favorably with beam energy due to improved overlap between the opening angles of the incoherent and superradiant emission \cite{Schaap2022PhotonElectrons}. The development of a superradiant Compton source would provide widespread availability to bright X-rays rivaling that of large scale facilities \cite{Graves2012Super}. 

\begin{figure} [b!]
    \centering
    \includegraphics[trim={0cm 0cm 0cm 0cm},width=0.5\textwidth]{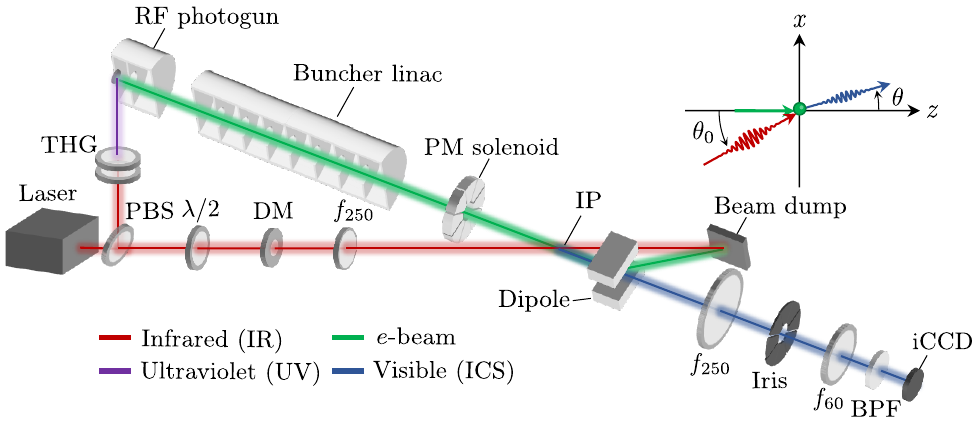}
    \caption{Schematic of the beamlines used in the experiment. See text for details.}
    \label{fig: 1}
\end{figure}

Based on this, a multitude of ICS sources \cite{Debus2010Traveling-waveSources,Dopp2023All-opticalAngles, Schaap2024IFEL}, including laser driven free electron lasers \cite{Lawler2013NearlyX-rays,Steiniger2014OpticalThomson-Scattering}, have been proposed in recent years involving such an overtaking scattering geometry. Despite the wide interest, the properties of shallow angle ICS have remained largely unexplored: the smallest crossing angle studied experimentally until date is 90 degrees \cite{Schoenlein1996FemtosecondMaterials,Leemans1996X-RayScattering,Taira2023}. In this Letter, we report the observation of Inverse Compton Scattering of visible light in such an overtaking geometry where a 4.7 MeV electron beam crossed a 780 nm laser at 5.8 $\deg$ angle. The visible spectral range of the emitted radiation allows us to leverage very sensitive detectors for extensive characterization of the interaction. The resulting data is found to be in good agreement with our theoretical predictions and numerical simulations. Moreover, due to the particular scattering geometry, differently than head-on scattering, polarization plays an important role in the interaction and in-plane ($p$) and out-of-plane ($s$) polarization yield different results. Importantly we observe, consistent with the predictions, a suppression of the emission for $p$-polarization at condition equivalent to relativistic Brewster reflection.

The experiment was carried out at the UCLA Pegasus photoinjector \cite{Alesini2015}. A schematic of the experimental geometry showing the shallow angle interaction and the diagnostic used to spectrally and angularly resolve the visible light radiation is shown in Fig. 1. A 200~pC electron bunch is emitted by illuminating a high quantum efficiency NaSkB photocathode \cite{Maxson2015} in a S-band 1.6 cell RF gun with a frequency tripled Ti:Sa laser system providing 260~nm 100~fs FWHM UV pulses. The gun is operated at 20$^{\rm o}$ phase and acceleration gradient of 65~MV/m and is followed by a buncher linac tuned at -80$^{\rm o}$ from crest with a gradient of 17~MV/m to compress the 4.7~MeV beam to 300~fs rms at the interaction point (IP) with the laser. A radially magnetized permanent magnet (PM) solenoid \cite{xu2025focusingrelativisticelectronbeams} located 8 cm upstream of the IP strongly focuses the electrons to a waist of about 150 $\mu$m at the interaction point. A PM dipole is used to separate the $e$-beam from the Compton radiation and minimize the radiation noise on the camera.


The infrared (IR) line that delivers the 780~nm laser pulse with 6~mJ pulse energy and 100~fs FWHM pulse length at the interaction point is first split by a polarizing beam splitter (PBS) from the same laser line that drives the photoinjector. Downstream of the PBS, the polarization angle $\phi_0$ of the linear polarized pulse is controlled by a half waveplate. The laser pulse is then focused by a lens ($f = 250$ mm) to a waist of 100 $\mu$m in the IP corresponding to a peak intensity of $3.5 \times 10^{14}$ W/cm$^2$. Upstream the lens we inserted a deformable mirror to compensate for astigmatism induced in the transport of the laser beam and that allowed live-adjustment to the laser spatial profile at the IP.

The Compton diagnostic line is a two lens point imaging system with a demagnification of $24\%$ to increase sensitivity to low photon counts collected on the 12.8 $\mu$m pixels of the intensified (i)CCD detector (Princeton Instruments PI-MAX 4). The detection system response was calibrated in-situ using photons from optical transition radiation (OTR) from a metal foil (see Supplemental Material (SM) \cite{SM}). A motorized iris is placed in the back-focal plane of the objective lens to angularly resolve the radiation distribution. To filter out any background laser light, we use a strong (OD $ > 15$) band pass filter (BPF) transmissive only in the 400-720 nm window.


As the spatiotemporal window for Compton scattering is very small, we use intermediate diagnostics to assist in alignment. For the spatial overlap a removable lanex screen was used, allowing for transverse profiling of both the laser and  $e$-beam at the IP. Temporal synchronization is obtained adjusting a linear delay stage while monitoring the relative time-of-arrival of the bunch with respect to the laser pulse by electric-optic sampling (EOS) of the $e$-beam Coulomb field in a birefringent ZnTe crystal (see SM \cite{SM}). 



\begin{figure} [b!]
    \centering
    \includegraphics[trim={0cm 0cm 0cm 0cm},width=0.5\textwidth]{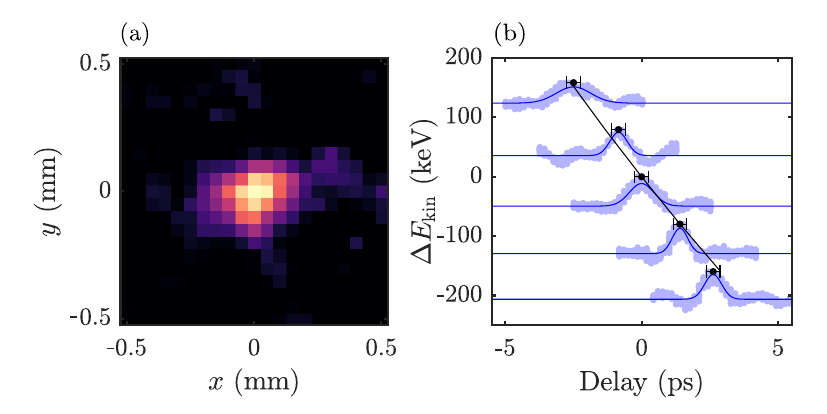}
    \caption{(a) Intensity measured on the iCCD averaged over 85 shots. (b) Energy of the beam relative to the nominal 4.7 MeV as function of arrival time at the IP represented by the black data points. The blue scattered data is the photon yield for each time delay.}
    \label{fig: 2}
\end{figure}

The measured intensity distribution from the resulting Compton scattering after overlapping in space and time the electron and laser beams is shown in Fig. \ref{fig: 2}(a). The rms source size is $88\times108$ $\mu\textrm{m}^2$, smaller than the measured $e$-beam size on the lanex and OTR screens, consistent with expectation from the convolution of the laser and electron distributions at the IP. The integrated yield captured within the maximal collection angle of $\gamma\theta_c = 0.8$ was typically 300 photons. 
Using additional low-pass filters with cutoff edges $\lambda_{\rm rad} <450$ nm and $\lambda_{\rm rad} <550$ nm respectively, we verified the emission spectrum to be consistent with the calculated on-axis radiation wavelength $\lambda_{\rm rad} = 414$ nm.


To experimentally demonstrate that the measured radiation is prompt with respect to the incident laser field, the $e$-beam energy was modified by changing the phase of the buncher linac. Since the time of arrival of the $e$-beam shifts due to the resulting velocity change, the delay line of the laser pulse was scanned for each energy. In Fig. \ref{fig: 2}(b) the energy of the $e$-beam is set out against the delay at which the Compton scattering time traces maximize, indicating optimal overlap. The time traces, shown by the blue scattered data in the figure, have been denoised using Wiener deconvolution due to the time-of-arrival jitter of the RF compressed $e$-beam.  (see SM \cite{SM}). The measured delays match well with the estimated 
time of arrival change given by $\Delta t_{\rm toa} = \Delta\gamma L/(\beta^2\gamma^3)$, with $L=197$ cm the distance between the exit of the linac and the IP, as indicated by the solid black curve in the figure. Furthermore, the rms duration of the Compton signals, found by fitting a Gaussian distribution to the time traces (solid blue curves), agree well with the EOS length, further confirming the promptness of the scattering. 

\begin{figure} [t!]
    \centering
    \includegraphics[trim={0cm 0cm 0cm 0cm},width=0.5\textwidth]{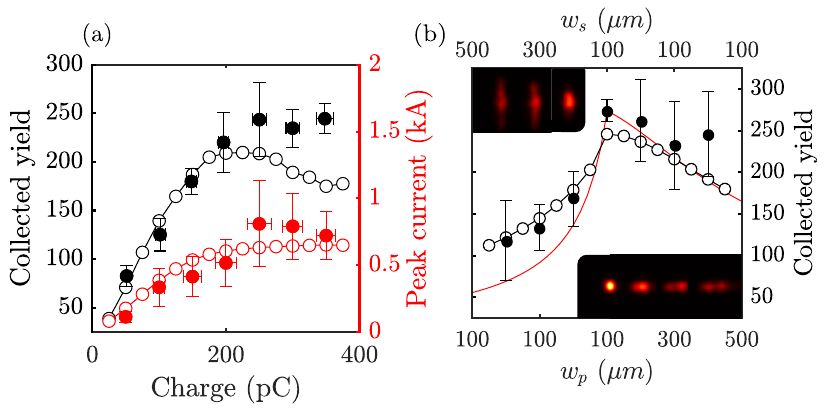}
    \caption{(a) Photon yield in black and peak current in red versus bunch charge as  (filled) measured and  (open) simulated. (b) Photon yield versus $s$ and $p$ laser waist size as (filled) measured and (open) simulated. The red solid curve represents the analytical prediction given by Eq. (\ref{eq: waist scaling}), which is normalized to the data at $w_{(s,p)} = 100$ $\mu$m. The insets in (b) show the transverse intensity distribution of the laser in the IP.}
    \label{fig: 3}
\end{figure}

The charge dependence of the photon yield was then investigated. The black dots in Fig. \ref{fig: 3}(a) display the collected yield for different charges as measured non-destructively with a integrating current transformer. Up until a beam charge of 150 pC the yield scales roughly linear as in the usual case for head-on scattering. However, at higher charges the yield starts to saturate. This results from space charge forces that oppose compression of the $e$-beam limiting the final peak current during interaction with the co-propagating laser pulse, shown by the red dots in Fig. \ref{fig: 3}(a). The measured peak current here was inferred independently using EOS. Furthermore,  the measurements are in good agreement with numerical calculations, given by the white dots, based on particle tracking performed with the general particle tracer (GPT) code combined with post-process radiation calculations based on classical Lieneard-Wiechert fields (see SM \cite{SM}). The deviation between the numerical and measured yield post saturation is potentially caused by clipping due to space charge dominated beam blow-up at {\it e.g.}, the edges of the PM solenoid. 

In another set of measurements we changed the laser pulse shape in the IP by imparting aberrations to the laser pulse with the deformable mirror. In particular, a non-zero astigmatism was applied to produce a wider spot in ($p$) and out ($s$) of the plane of interaction, independently, while the pulse energy remains constant. The insets in Fig. \ref{fig: 3}(b) show the intensity distributions of the laser in the interaction point as measured on the lanex screen. The resulting collected yield for each laser spot, in very good agreement with numerical simulations displayed in Fig \ref{fig: 3}(b), reveals a scaling difference between the two waists $w_{s,p}$, despite equivalent reduction in laser intensity. The asymmetry arises in a side-scattering geometry since the interaction length, given by $t_{\rm int} = t_k t_p(t_k^2 + t_p^2)^{-1/2}$, where $t_k = \sigma_t/(1 - \beta \cos \theta_0)$ is the longitudinal transit time, $\sigma_t$ is the rms laser pulse length and $t_p = w_p/(2^{1/2}c\beta\sin\theta_0)$ is the transverse transit time, is affected by the waist $w_p$ in the $p$-direction \cite{Schaap2024IFEL}. The Compton yield, proportional to the product of the intensity and the interaction time, therefore scales as
\begin{equation}  \label{eq: waist scaling}
    N_{\rm ph} \propto \frac{1}{w_s} \frac{1}{\sqrt{2\beta^2c^2\theta_0^2 t_k^2  + w_p^2}},
\end{equation}
where we expanded for shallow crossing angle and $t_k = 3.9$ ps in our case. The first term in the square root is the distance traveled transversely by an electron through the laser pulse within the longitudinal overtaking time. The simple analytical model matches well with the measured trend as shown by the solid red curve in Fig \ref{fig: 3}(b) with minor deviations resulting from the finite extent of the electron beam. This scaling suggests control over radiation pulse length and bandwidth by laser focusing in the $p$-direction without a significant drop in flux. Pulse front tilting of the laser by $\sim\!\theta_0/2$ would completely remove the limiting contribution of the longitudinal transit time such that $t_{\rm int} = t_p$, enabling an even larger tunability range and avoiding restrictions on the photon yield imposed by laser diffraction \cite{Debus2010Traveling-waveSources}.

So far, the measured dependencies of the yield are mainly affected by the specific geometric constraints in oblique Compton scattering. Now, we turn to study the effect of the incident laser polarization, which under a shallow angle results in surprising changes of the radiation properties. In particular, we show that coupling to the electron beam under a specific crossing angle $\theta_\beta$, which we refer to as the relativistic Brewster angle, orients the dipole moment of the electrons from transverse for $s$-polarization to longitudinal for $p$-polarization. This results in suppressed emission along the electron beam axis for the latter case, an effect completely analogous to the well-known Brewster reflection in optics.


\begin{figure} [t!]
    \centering
    \includegraphics[trim={0cm 0cm 0cm 0cm},width=0.35\textwidth]{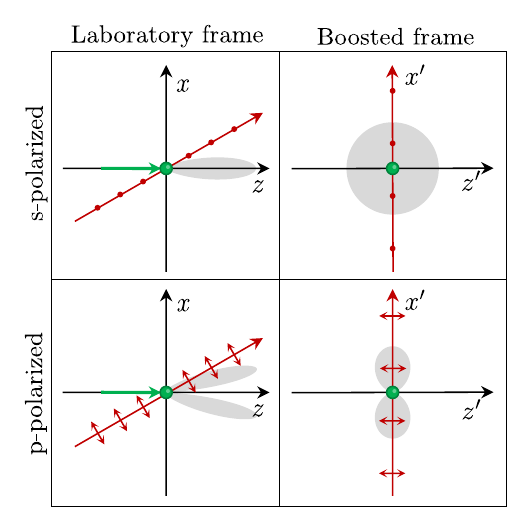}
    \caption{Schematic of Compton scattering with laser (red) under the relativistic Brewster angle with corresponding differential cross-section (grey) in boosted and laboratory frame.}
    \label{fig: 4}
\end{figure}

Qualitatively, it is insightful to discuss what happens in the instantaneous rest frame of the electrons characterized by a coordinate system ($x',y',z'$), where the electrons are wiggled by the Lorentz-transformed laser wave and emit dipole radiation. In this frame, the angle of incidence transforms as $\tan\theta_0' = \sin\theta_0/[\gamma(\beta -\cos\theta_0)]$ \cite{Einstein1905ZurKorper}. We define the relativistic Brewster angle as the crossing angle which in the rest frame is normal to the electron beam axis ($\theta_0' = \pi/2$). This occurs at $\theta_0 =   \arccos \beta \equiv \theta_\beta$, which for relativistic beam can be approximated by $\theta_\beta = \gamma^{-1}$. 
Figure \ref{fig: 4} depicts the radiation pattern resulting from $s$- and $p$-polarization in the laboratory and boosted frame for the relativistic Brewster angle case. For an s-polarized laser pulse, the resulting dipole radiation in the boosted frame is concentrated in the $x'z'$-plane and no emission in the $y'$-direction. In the laboratory frame this results in a radiation pattern with an opening angle of $\theta_\beta$ along the polarization direction and a maximum along the $e$-beam axis as in the traditional ICS case. However, for $p$-polarization under this angle the induced dipole moment is exactly parallel to the $z'$ axis leading to a dipole radiation in the transverse directions and no energy radiated along the beam axis of the boosted frame. In the laboratory frame this corresponds to an annular radiation pattern with vanishing on-axis emission. 

Quantitatively, the angular emission pattern in the laboratory frame can be calculated using the differential cross-section. For shallow crossing and small emission angles, $\theta_{(0)}\ll 1$ in the relativistic limit, $\gamma \gg 1$, the ICS differential cross-section can be written as \cite{SM}:
\begin{widetext}
\begin{equation} \label{Eq: differential cross-section}
\frac{d \sigma}{d \Omega} = \frac{4 \gamma^2 r_e^2}{(1 + \gamma^2\theta^2)^2}\left\{1 - 4\left[\frac{ \gamma\theta\sin(\phi -\phi_0) + \gamma^3\theta_0^2 \theta \sin(\phi + \phi_0) + (\gamma^2\theta^2 -1 )\gamma\theta_0\sin\phi_0 }{(1 + \gamma^2\theta^2) ( 1 + \gamma^2\theta_0^2)}  \right]^2\right\},
\end{equation}
\end{widetext}
where $r_e$ is the classical electron radius. It is clear that for $s$-polarization ($\phi_0 = 0$) this expression simplifies to the familiar form $d\sigma/d\Omega  = 4\gamma^2r_e^2/X^2[1 - 4(X-1)\sin^2\phi/X^2]$ where $X = 1 + \gamma^2\theta^2$, which is independent of crossing angle $\theta_0$. For $p$-polarization ($\phi_0 = \pi/2$) at the Brewster angle, the differential cross-section condenses to an unusual form given by  $d\sigma/d\Omega  = 4\gamma^2r_e^2(X-1)/X^3$, which maximizes at scattering angle $\theta_{\rm max}=  \pm 3^{-1/2}\gamma^{-1}$. Another interesting case that exemplifies the analogy to Brewster reflection is the differential cross-section along the main scattering axis ($\theta=0)$ given by  $d\sigma/d\Omega(0) = 4\gamma^2 r_e^2[1 - 4(X_0-1)\sin^2\phi_p/X_0^2]$ where $X_0 = 1 + \gamma^2\theta_0^2$, which vanishes only for the specific case of  $\theta_0 = \theta_\beta$ for $p$-polarization. The expression also indicates that for most cases, other than close to the Brewster angle, the intrinsic scattering efficiency along the beam axis is not affected by a shallow angle. 

\begin{figure} [h!]
    \centering
    \includegraphics[trim={0cm 0cm 0cm 0cm},width=0.45\textwidth]{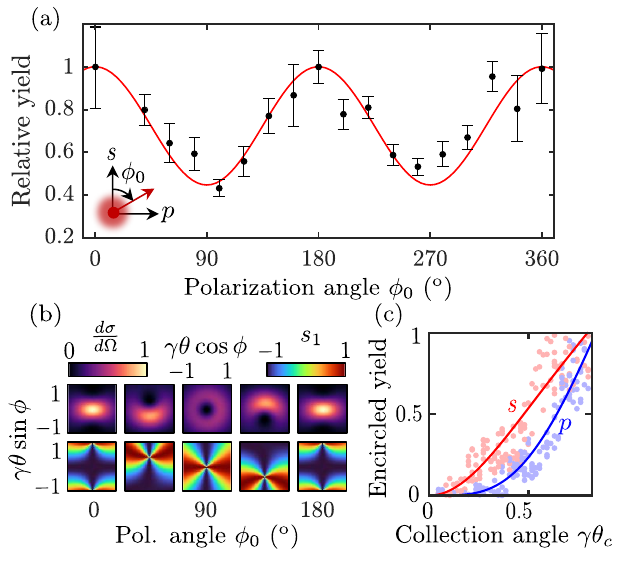}
    \caption{(a) Collected yield for polarization angle at collection angle of $\gamma\theta_c = 0.5$. The red curve, represented by Eq. (\ref{eq: collected yield}) fits well with the data. (b) Differential cross-section as given by Eq. (\ref{Eq: differential cross-section}) and degree of linear polarization $s_1 = (|E_x|^2 -|E_y|^2)/(|E_x|^2 +|E_y|^2)$ with $E_{(x,y)}$ the electric field components in the $x$ and $y$-direction, respectively. (c) Photon yield as function of collected angle for (red) $s$- and (blue) $p$-polarization. The solid curves are given by Eq. (\ref{eq: collected yield}). }
    \label{fig: 5}
\end{figure}

Given that in the experiment the crossing angle is $\theta_0 = 0.94 \theta_\beta$, we expect that the polarization will significantly affect the produced radiation. However, since the source plane is imaged, we can only angularly resolve the number of photons for a given collection angle by changing the iris opening. The collected yield is proportional to the integral $N_{\rm ph} \propto \int_{0}^{\theta_c} \int_{0}^{2\pi} d\sigma/d\Omega \sin\theta d\theta d\phi$, which can be evaluated analytically as:
\begin{equation} \label{eq: collected yield}
N_{\rm ph }\!\propto\! \frac{2\gamma^4\theta_c^4}{(1\! +\! \gamma^2\theta_c^2)^2}\! +\! \left[\ln\left(1\!+\!\gamma^2\theta_c^2\right)  - \frac{3\gamma^4\theta_c^4}{(1\! +\! \gamma^2\theta_c^2)^2}\right]\!\cos^2\phi_0,
\end{equation}
where we assumed $\theta_0 = \theta_\beta$. This expression exhibits a $\cos^2 \phi_0$ behavior, with a maximum for $s$-polarization, a minimum at $p$-polarization and with a modulation depth that is determined by the collection angle. In the final set of measurements we probe the polarization dependent yield given by the Eq. (\ref{eq: collected yield}).

Figure \ref{fig: 5}(a) displays the measured collected yield for $\gamma \theta_c = 0.5$ by the black dots. The data agrees very well with the theoretical prediction both in polarization dependence and in modulation depth. Several corresponding differential cross sections given by Eq. (\ref{Eq: differential cross-section}) are shown below in Fig. \ref{fig: 5}(b) together with the calculated polarization of Compton scattering, which changes from linear, as is usual for ICS from a linear polarized laser pulse \cite{Petrillo2015Pol}, to radial polarization. Last, we measured the yield for $p$- and $s$-polarization as a function of collection angles as shown in Fig. \ref{fig: 5}(c). These results match the theoretical prediction highlighting the change from the conventional spatial radiation pattern in ICS to an annular pattern for $p$-polarization close to the Brewster angle. 

To summarize, we have observed inverse Compton scattering of visible light from a relativistic $e$-beam crossed by a co-propagating laser pulse under a shallow angle. We measured scattering efficiency for varying charge and laser spot size, showing significantly different scaling than in conventional ICS sources. Furthermore we find that the Compton emission from a relativistic electron beam in this configuration has a strong polarization dependence which can be interpreted as relativistically analogous to Fresnel reflection close to the Brewster angle. The experimental results show good agreement with start-to-end simulations and analytical predictions. These results greatly advance the development of compact radiation sources based on Compton scattering enabling 
spectral tunability and brightness increase from the use of higher energy $e$-beams. Notably, this shallow angle geometry is expected to play an important role in unfolding the potential of operating in the superradiant regime for coherent Compton emission.

\begin{acknowledgments}

We thank Jared Maxson, Chad Pennington and David Garcia for providing the NaKSb cathode used in the experiment, Tianzhe Xu for providing the permanent magnet solenoid and Coen Sweers for insightful discussions.  This research is supported by DOE grant No. DE-SC0009914. This work was also partially supported by the National Science Foundation under Grant No. PHY-1549132.
\end{acknowledgments}

\bibliography{references}

\end{document}